\documentclass[final,5p,times,twocolumn]{elsarticle}

\usepackage{amssymb}

\journal{Physics Letters B}
\usepackage{xcolor}
\usepackage{amsmath}

\def\ba{\begin{array}}
\def\ea{\end{array}}
\def\bea{\begin{eqnarray}}
\def\eea{\end{eqnarray}}
\def\beq{\begin{equation}}
\def\eeq{\end{equation}}
\def\ben{\begin{enumerate}}
\def\een{\end{enumerate}}

\def\brr{\begin{array}}
\def\err{\end{array}}

\def\rL{r_\Lambda}
\def\chiSW{\chi_{*}}

\def\LCDM{$\Lambda$CDM }
\def\dA{{d\Omega}}

\def\black{\color{black}}

\def\in{{\bf{-}}}
\def\out{{\bf{+}}}

\begin{document}

\begin{frontmatter}



\title{What Moves the Heavens Above?}


\author[inst1,inst2]{Enrique Gazta\~naga}
\author[inst1,inst2]{Benjamin Camacho-Quevedo}

\affiliation[inst1]{organization={Institute of Space Sciences (ICE,CSIC)},
            addressline={}, 
            city={Bellaterra},
            postcode={08193}, 
            state={Barcelona},
            country={Spain}}
\affiliation[inst2]{organization={Institut d'Estudis Espacials de Catalunya (IEEC)},
            addressline={}, 
            city={Barcelona},
            postcode={08034}, 
            state={Barcelona},
            country={Spain}}            

\begin{abstract}
   The standard cosmological model (\LCDM) assumes that everything started in a singular Big Bang out of Cosmic Inflation, a mysterious form of modern Aether (the inflaton).
   Here we look for direct observational evidence for such  beginning in
   two  recent measurements: 1) cosmic acceleration, something \LCDM attributes to Dark Energy (DE), 2) discordant measurements for $H_0$ and anomalies in the CMB.
   We find here that observed variations in $H_0$ correspond to large metric perturbations that are not consistent with the simplest models of Inflation or DE in the \LCDM paradigm. Together, these observations indicate instead that cosmic expansion could originate  from a simple gravitational collapse and bounce. We conjecture that such bounce is trigger by neutron degeneracy at GeV energies. 
   This new paradigm explains the heavens above using only the known laws of Physics, without any new Aether, DE or Inflation.
\end{abstract}




\end{frontmatter}


\section{Introduction}
\label{S:1}

 Aristotle proposed that Aether moved the heavenly spheres of stars and planets. It took two millennia for the likes of  Copernicus and Newton to realize that such Aether was a combination of Gravity and Earth's rotation (also due to Gravity). Yet, according to the current lore, cosmic expansion can not be explained by Gravity alone. 
The standard cosmological model \cite{Dodelson}, also called $\Lambda$CDM, assumes that our Universe began in a hot Big Bang from Cosmic Inflation. 
The assumption that
the expansion is driven by gravity all the way back to the beginning ($\tau=0$) raises the well-known horizon problem. This is a profound problem that challenges our understanding of Cosmology or Gravity.
The \LCDM model partially solves this problem by introducing Cosmic Inflation, an ad hoc theory that is very hard to test because it is based on fields and energies ($10^{15}-10^{19}$GeV) that are far beyond what we can ever access with experiments. There is no evidence that our Big Bang ever reached such large energies.  
The detailed mechanism responsible for inflation  or what was before are still unknown \cite{Dyson,PenroseEntropy,Brandenberger,Steinhardt}. 
More speculative alternatives to Inflation exist outside the known laws of Physics, within Quantum Gravity and the Brane World (see, e.g., \cite{2005PhLB..614..125D} and references therein).
Inflation also provides a prediction for the initial conditions of the observed large-scale structures, but such predictions are too generic and given in terms of free parameters. The simplest models of Inflation predict adiabatic scale invariant fluctuation in general agreement with current observations \cite{Dodelson}.

Additional conceptual problems of \LCDM are the need to include Dark Matter and Dark Energy, for which we have no direct evidence or fundamental understanding.
With these fixes, the \LCDM model seems to provide a very successful framework to understand most cosmological and astrophysical observations by fitting a few free cosmological parameters. However, recent observations of anomalies in these fits (see \cite{2022arXiv220306142A} for an extensive summary) show discrepancies with the \LCDM predictions that are hard to explain.
Here, we will focus on variations in the expansion rate today, $H_0$, over super-horizon scales across the sky. But similar arguments apply to other cosmological parameters (for example the $\sigma_8$ tension). We will assume that such variations are due to perturbations around a uniform model and show that they are inconsistent with the \LCDM predictions. We will end by proposing an alternative explanation.

\section{Anomalies in \LCDM}
\label{sec:super-horizon}

The \LCDM model assumes that the whole Universe can be modeled as an infinite homogeneous and isotropic background space, given by the Friedmann–Lemaitre–Robertson–Walker (FLRW) metric, where the physical radial distance is given by $dr=a d\chi$ in terms of co-moving coordinates $d\chi$. The adimensional scale factor, $a=a(\tau)$, gives the expansion as a function of co-moving time $\tau$. In the physical (or rest) frame, $r=a\chi$, which implies $\dot{r}=H r$, where $H \equiv \dot{a}/a$, for a comoving observer. The Hubble Horizon is
$r_H \equiv c/H$ (we will use units of speed of light $c=1$), so that $r>r_H$ implies $\dot{r}>1$ and corresponds to super-horizon scales. For a perfect fluid with density $\rho$ and pressure $p$, the solution to the field equations in a flat space is well known:
\beq
 H^2  =  \frac{8\pi G}{3} \rho =  H_0^2  \left[ \Omega_m a^{-3} + \Omega_R a^{-4} + \Omega_\Lambda \right]
\label{eq:Hubble}
\eeq
with $\Omega_{X} \equiv \rho_X/\rho_c$ and $\rho_c \equiv 3 H_0^2/(8\pi G)$. The current ($a=1$) matter density is given by $\Omega_m$, while radiation is given by $\Omega_R$. The effective cosmological constant term, $\Omega_\Lambda$, derives from $\rho_\Lambda  \equiv \Lambda/(8\pi G)$.
At any time, the expansion/collapse rate $H^2$ is given by
$\rho$. Energy--mass conservation requires that
$\rho \propto a^{-3(1+\omega)}$, where $\omega=p/\rho$ is the equation of state of the different components:
$\omega=0$ for pressureless matter (or dust), $\omega=1/3$ for radiation, and $\omega=-1$ for $\rho_\Lambda$.

\begin{figure}
\centering\includegraphics[width=.85\linewidth]{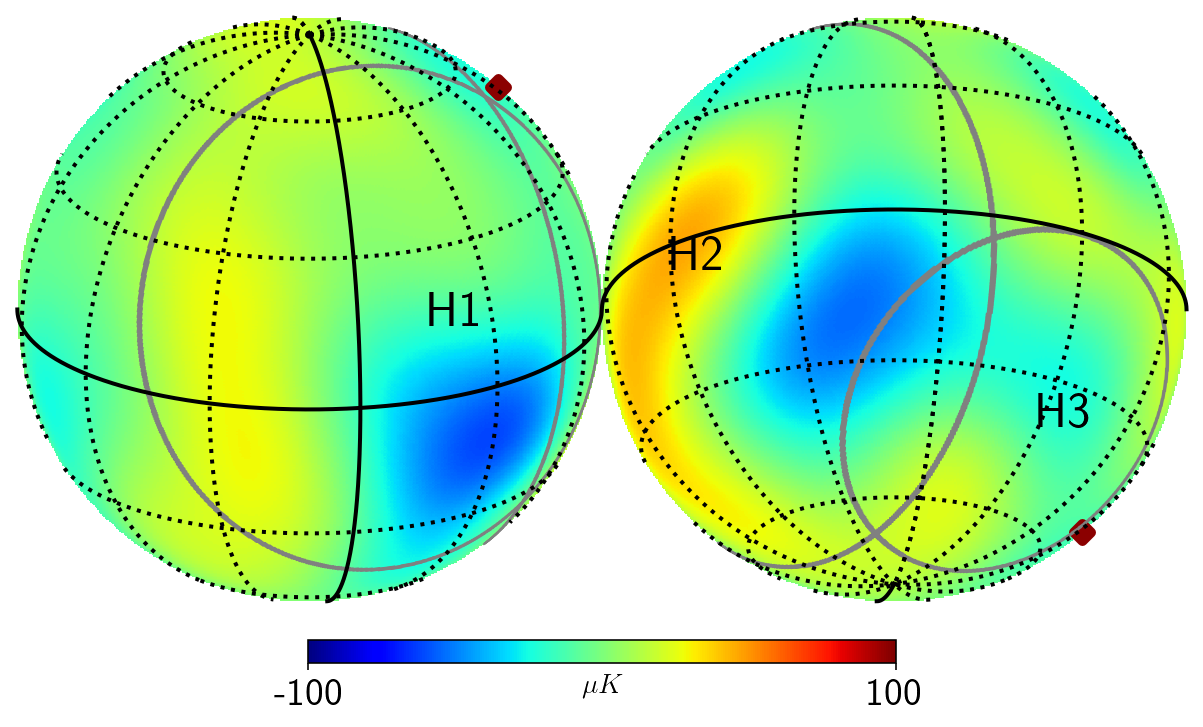}
\centering\includegraphics[width=.9\linewidth]{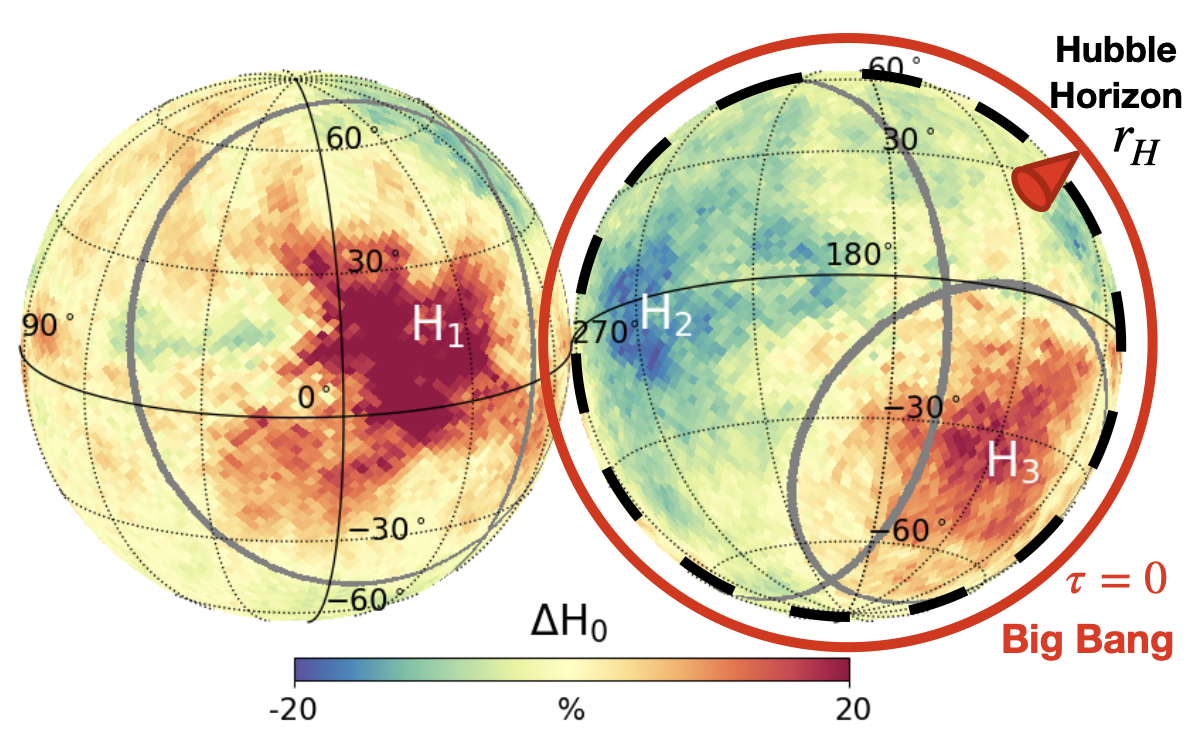}
\caption{Top panels show two view angles to the full sky with 
$\delta T$ in the observed CMB temperature $T$ (Planck SMICA map) smoothed with a 30 deg Gaussian. Bottom panel shows the relative variation in $H_0$ obtained from a fit to the small-scale $\delta_T$ power spectrum $C_\ell$ estimated over large regions of the sky \cite{FG20}.
These are displayed on the surface of a sphere whose radius $\chi_*$ is the distance traveled by the CMB light to reach us. The red circle represents the Big Bang surface ($\tau=0$). The horizon (small red cone) is the distance traveled by light between $\tau=0$ and the CMB. Large gray circles on the CMB surface are circular super-horizon boundaries (labeled $H_1$, $H_2$ and $H_3$), where $\Delta H_0$ vanish.}
\label{fig:CMBhorizons}
\end{figure}

\subsection{Cosmic Acceleration}

Cosmic acceleration is defined as $q \equiv  (\ddot{a}/ a) H^{-2}= -\frac{1}{2} (1+3\omega)$. For regular matter, we have $\omega>0$, so we expect the expansion to decelerate ($q<0$) because of gravity. However, the latest concordant measurements from a Type Ia supernovae (SNe), galaxy clustering, and CMB all agree with $\omega = -1.03 \pm 0.03$ \cite{DES2021} or $q \simeq 1$ in the future.

Note how $q=1$ (or $\omega=-1$) implies that $\rho$ and $H$ become constant, $H=H_\Lambda  \equiv 1/r_\Lambda$. Constant velocity is equivalent to no velocity in the rest frame, were all structures become super-horizon and freeze.
In the physical (or rest) frame, such expansion is not accelerating but is asymptotically static.
This is important because it shows that we are trapped inside a future Event Horizon. This frame duality can be understood as a Lorentz contraction $\gamma=1/\sqrt{1-\dot{r}^2}$, where $\dot{r}= Hr$. An observer in the rest frame, not moving with the fluid, sees the moving fluid element $a d\chi$ contracted by the Lorentz factor $\gamma$.
Therefore, the FLRW metric becomes de-Sitter like:
\beq
a^2 d^2\chi = \gamma^2  d^2r =  \frac{d^2r}{1-r^2/r_H^2} \equiv \frac{d^2r}{1+2\Phi},
\label{eq:dS}
\eeq
where $\Phi$ is the gravitational potential, which can also be interpreted as a metric perturbation. This radial element corresponds to the metric of a hypersphere of radius $r_H$ that expands towards a constant event horizon $\rL$. In this limit, the FLRW metric reproduces the static de-Sitter metric: $2\Phi=-r^2 H_\Lambda^2$, where the expansion becomes static (see also \cite{Mitra2012} and the Appendix here).

\subsection{Super-horizon perturbations}

Structures larger than the Hubble Horizon $r_H$ are not in causal contact because the time that a perturbation takes to travel such distance is larger than the expansion time. As $r_H$ increases with $\tau$, the structures we observe today were not in causal contact in the early Universe (e.g., in the CMB). This is the horizon problem. In the \LCDM model, the problem is solved by Cosmic Inflation \cite{Starobinski1979,Guth1981,Linde1982,Albrecht1982}, a period of exponential expansion that happened right at the beginning of time.
Inflation solves the horizon problem but leaves the universe empty. We need a mechanism to stop Inflation and to create the matter and radiation that we observe today.

\begin{figure*}
\begin{center}
\includegraphics[width=0.49\textwidth]{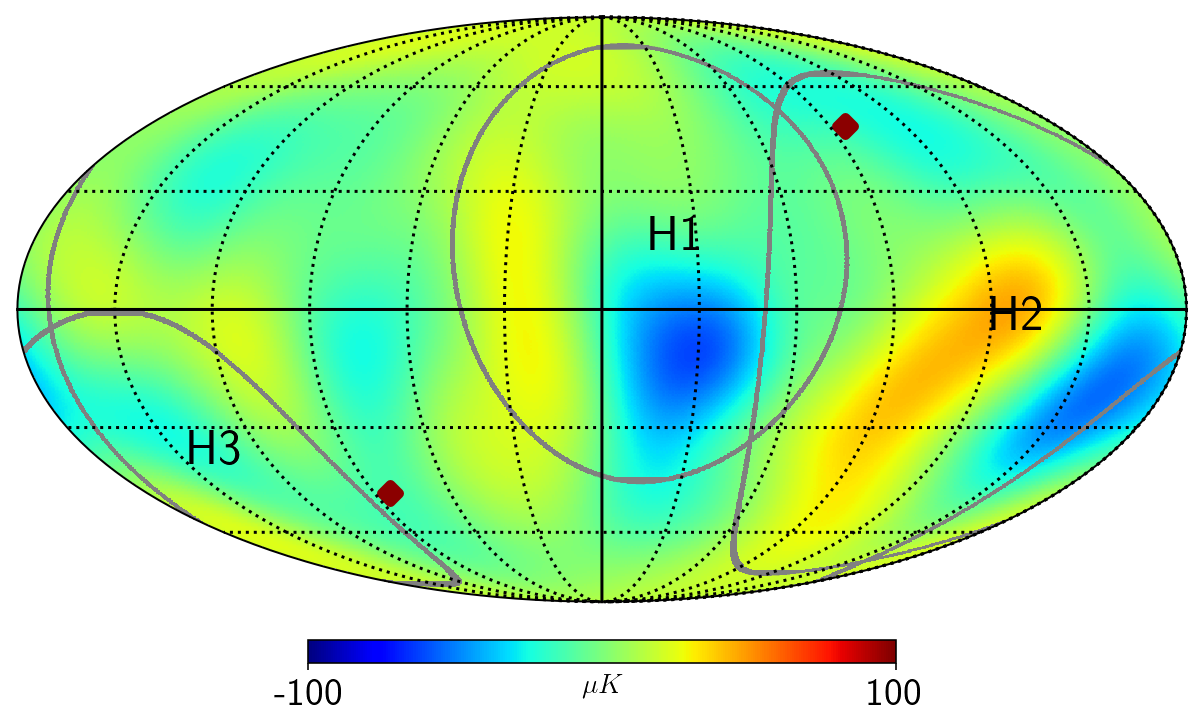}
\includegraphics[width=0.49\textwidth]{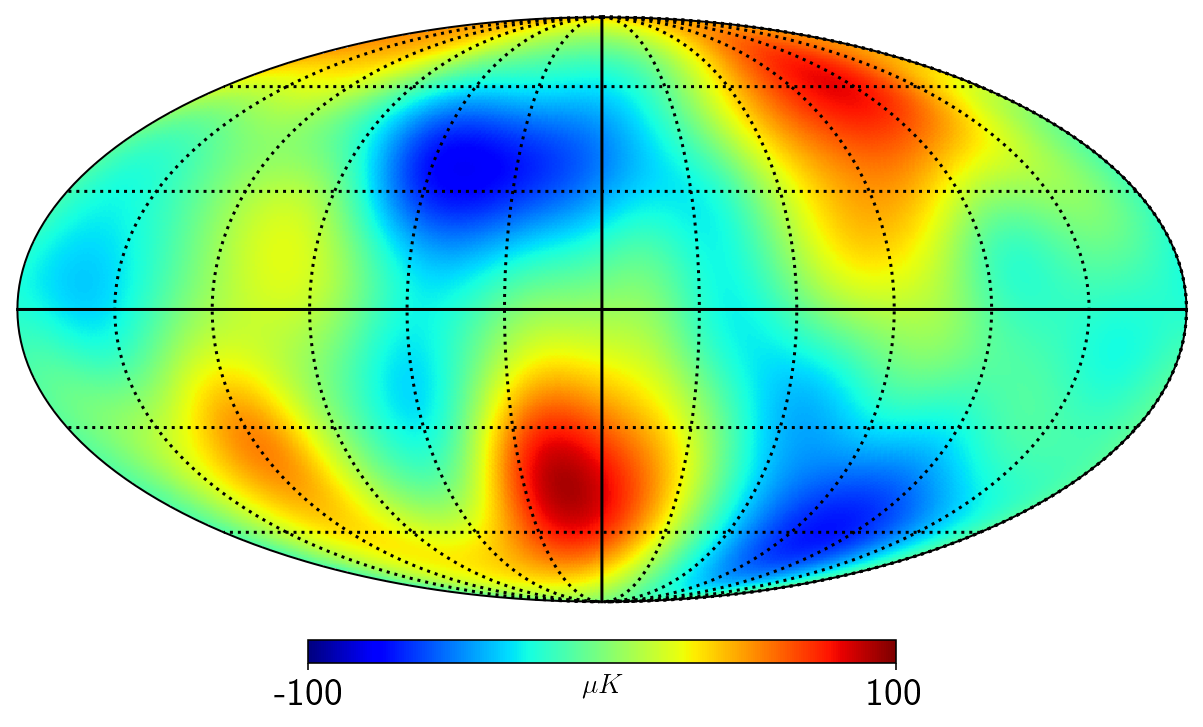}\caption{Comparison of CMB temperature $\delta T$ maps from data (Planck SMICA map, left) and a simulation of the best fit $\Lambda$CDM model (right), smoothed with a 30 deg. Gaussian radius. Both maps have very similar amplitude in small scale angular modes ($C_\ell$, not shown), but there is a lack of power at the largest super-horizon scales ($\theta >60$ deg.) in the real data. We also show the $H_0$ horizons displayed in Fig.\ref{fig:CMBhorizons} as gray circles over the Planck map. The CMB dipole direction, shown as two red diamonds, does not seem to be related to the super-horizon perturbations.}
\label{fig:CMB-maps}
\end{center}
\end{figure*}

If we could see the light from the Big Bang ($\tau=0$), it would come from a very distant spherical shell in the sky (red circle in the bottom right of Fig.\ref{fig:CMBhorizons}). The furthest we can actually see is the CMB shell (dashed circle $a_* \simeq 10^{-3}$), which is quite close to $\tau=0$ ($a=0$). This means that $r_H \propto c\tau$ subtends a very small angle in the sky: $ \theta = r_H / (a_* \chi_*) \simeq 1$ deg., where $\chi_*$ is the co-moving angular diameter distance to the CMB. Larger scales are super-horizon. How is then possible that the CMB temperature is so uniform on larger scales (as shown by top panel in Fig.\ref{fig:CMBhorizons})?
Inflation can solve this puzzle, but a collapsing phase, before the Big Bang, can also do so.

In the simplest models of 
Inflation, the spectrum of primordial super-horizon perturbations is scale invariant and adiabatic, so we expect to see temperature and metric perturbations of equal size at large scales. However, there is an anomalous lack of the largest structures in the CMB sky temperature $T$ with respect to the predictions of Inflation (see, e.g., \cite{Gaztanaga2003,Schwarz2016}).
This is apparent in Fig. \ref{fig:CMB-maps}, which compares $\delta T$ in the CMB sky with a \LCDM simulation.
The CMB isotropy scale can be measured with the homogeneity index, $\mathcal{H}$, a fractal or Hausdorff dimension that is model-free and purely geometrical, independent of the amplitude of $\delta T$.
\cite{Benjamin} find evidence of homogeneity ($\mathcal{H}=0$) for scales larger than $\theta_{\mathcal{H}} = 65.9 \pm 9.2$ deg. on the CMB sky. This finding is at odds (with probability  $p < 10^{-5}$) with the predictions of Inflation.

A related anomaly is shown at the bottom of Fig. \ref{fig:CMBhorizons}. It displays a sky map of relative variations of $H_0$ from the best fit to the standard \LCDM temperature spectrum $C_\ell$ (for $32<\ell<2000$)
over large regions ($\theta >$ 30 deg.) around each position in the sky (see \cite{FG20} for details and \cite{2022arXiv220103799Y} for similar results).
The fit in each region agrees well with the predictions, but the fitted parameters vary across the sky. Here, we interpret such variations as super-horizon perturbations.

There is a characteristic cut-off scale where $\Delta H_0$ vanishes, which is shown by gray circles labeled $H_1, H_2, H_3$. The same horizons are found for different cosmological parameters. They correspond to a cut-off in super-horizon perturbations from the $\tau=0$ surface,
indicating that the primordial spectrum is not scale invariant as predicted by the simplest models of Inflation.

\begin{figure}
\centering\includegraphics[width=.8\linewidth]{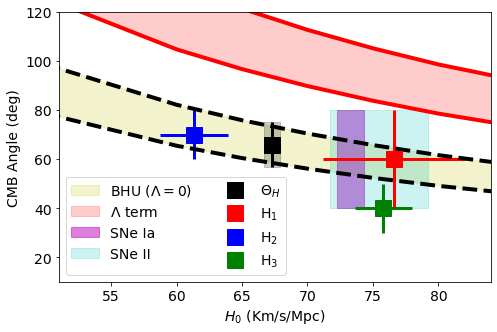}
\caption{The angular size at the CMB sky
 of the causal horizons in Fig.\ref{fig:CMBhorizons} (colored squares) and the homogeneity scale $\theta_H$ (black square) as a function of the mean value of $H_0$ in each region. The mean estimates for $H_0$ in Planck (around $\theta_H$), SNe Ia (pink) and SNe II (cyan) are shown as shaded 68\% confidence regions. The angle for SNe corresponds to the radial distance to the CMB projected in the CMB sky. Yellow filling shows the BHU prediction in Eq.\ref{eq:cut-off} with
$\Lambda=k=0$ during collapse. Red filling uses $\Omega_\Lambda=0.75 \pm 0.05$ instead.}
\label{fig:CMBhorizons2}
\end{figure}

In Fig.\ref{fig:CMBhorizons2}, we compare the angular size of $H_1, H_2, H_3$ with the homogeneity scale $\theta_{\mathcal{H}}$ as a function of the mean value (and dispersion) of $H_0$ measured in each region. The $\theta_{\mathcal{H}}$ measurement corresponds to the full sky and is therefore assigned to the global Planck fit for $H_0$ \cite{P18cosmo}. We can also place in the same plot the local type Ia SN measurement of $H_0$ from \cite{SHOES} and the one from type II SN from \cite{2022arXiv220308974D}.
Similar results are found using time delays in a lensed QSO \cite{2020MNRAS.498.1420W}.
For SNe, the angle is taken to be 60 degrees, as this is the angle in the CMB sky that corresponds to the radial separation $\chi_*$ between the CMB surface and the SNe measurements. The angular spread is taken from the largest variance in the other measurements.

Note that the $\theta_{\mathcal{H}}$ and $H_1, H_2, H_3$ measurements are independent from each other. The latter are obtained by fitting the small-scale $C_\ell$ power spectrum ($\le 1$ deg. ), while $\theta_{\mathcal{H}}$ is measured from the scaling of the correlations at much larger scales ($\ge 30$ deg.). The SNe results are corrected to the CMB frame, but this has a negligible impact on $H_0$ \cite{SHOES,2022arXiv220308974D}.

\begin{figure}
\centering\includegraphics[width=1.\linewidth]{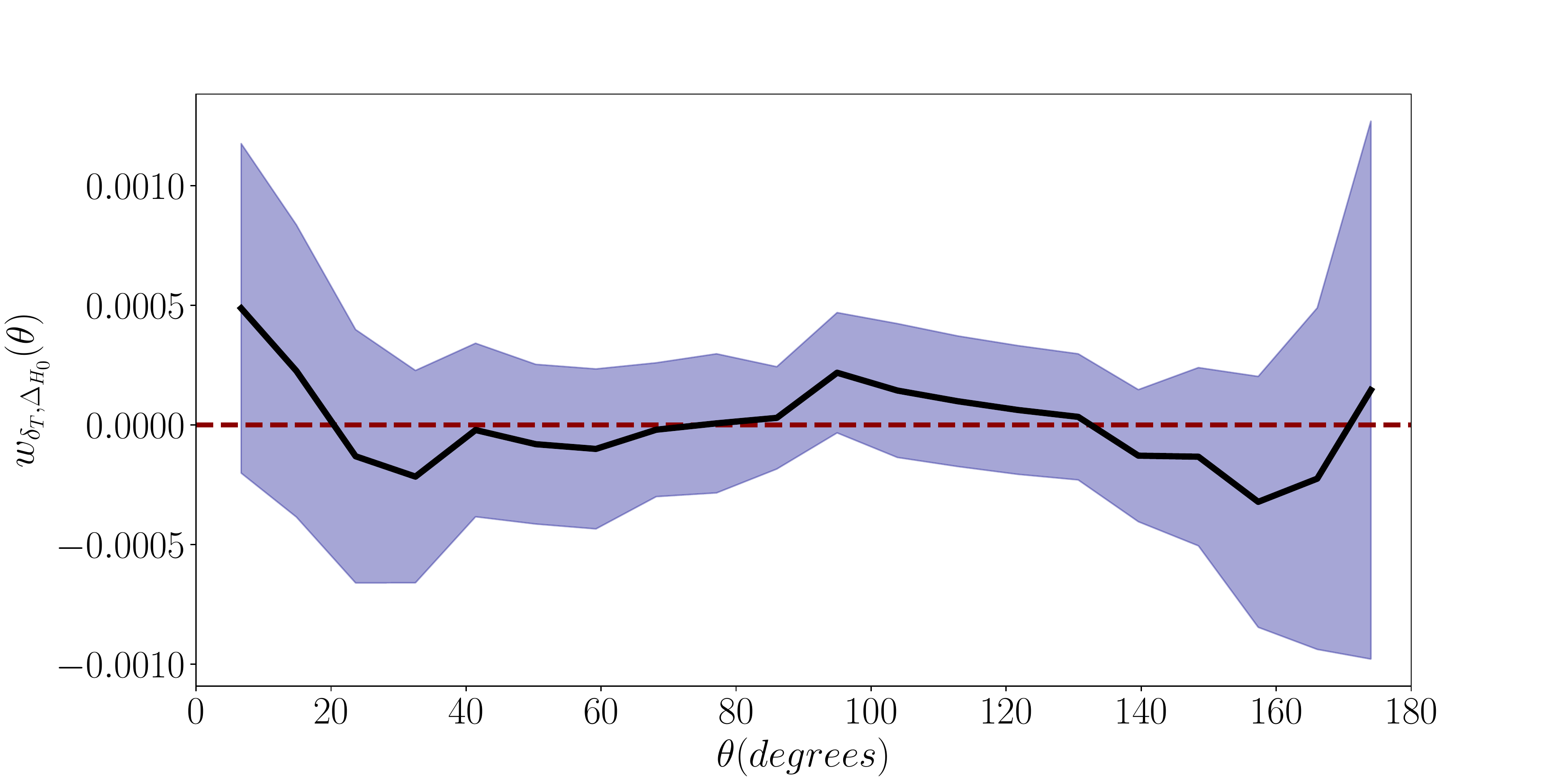}
\caption{
Angular 2-point cross-correlation $w(\theta)$ between the normalized temperature fluctuation maps $\delta T$ and the normalized $H_0$ fluctuations $\delta_{H_0}$ in Fig. \ref{fig:CMBhorizons}. Shaded region corresponds to Jack-Knife errors (see \cite{Benjamin} for implementation details).
The fact that the measured cross-correlation is consistent with zero shows that underlying fluctuations are not adiabatic.}
\label{fig:CMBhorizons3}
\end{figure}

Note how the variations of $H_0$ within the CMB are of similar size to the ones between CMB and SNe. 
This is what we expect for metric perturbations in Eq.\ref{eq:dS},
where $2\Phi=-r^2 H^2$, so that:
\beq
\delta_\Phi
\simeq 2 \frac{\Delta H_0}{H_0} =  \frac{\Delta \rho_0}{\rho_0} \simeq 0.2.
\label{eq:deltaH0}
\eeq
Large-scale metric perturbations do not evolve with time, so we expect these variations both in nearby and CMB observations, in agreement with Fig.\ref{fig:CMBhorizons2}.
However, the large values in Eq.\ref{eq:deltaH0} are at odds with the small amplitude ($\delta_T \simeq 10^{-5}$) and location of radiation fluctuations in the CMB sky, as shown by comparing the top and bottom panels in Fig.\ref{fig:CMBhorizons} 
and the null cross-correlation in 
 Fig.\ref{fig:CMBhorizons3}.
\black
This indicates that, on the largest super-horizon scales, perturbations are not adiabatic, as predicted by Inflation \cite{Dodelson}. 

\section{The black hole universe}

We next present an alternative to Inflation that could explain such anomalies. We call it the black hole (BH) universe (BHU, \cite{BHU1,BHU2,GaztaUniverse}), where cosmic expansion originates from the gravitational collapse of a dust cloud (from a large and very low density, almost Minkowski background). To understand this, we first notice that we live inside a BH, which is defined as an object of mass $M$ with escape velocity $\dot{r}>1$, which is equivalent to say that the object has a size $R \le r_S$, where:
\beq
r_S =2GM \simeq 2.9 \textrm{km} \frac{M}{M_{\odot}}
\simeq 1 \textrm{Kpc}  \frac{M}{10^{16} M_{\odot}}
\label{eq:BHdef}
\eeq
is the Schwarzschild (SW) radius or BH event horizon.
For an observer in flat space outside the BH, the mean density inside $r_S$ is always:
\beq
\rho_{BH} 
=\frac{3  r_S^{-2}}{8\pi G}
\simeq 10^{-2}  \left[\frac{M_{\odot}}{M}\right]^2 \, \frac{M_{\odot}}{\textrm{km}^{3}}  \, .
\label{eq:BHrho}
\eeq
This can be compared to the atomic nuclear saturation density:
\beq
\rho_{NS} \simeq 2 \times 10^{-4} \frac{M_{\odot}}{\textrm{km}^{3}} \,
\label{eq:NS}
\eeq
which corresponds to the density of heavy nuclei. 
The latter results from the Pauli Exclusion Principle applied to neutrons and protons. For a neutron star (NS) with $M \simeq 7M_{\odot}$, we have $\rho_{BH}=\rho_{NS}$. This explains why we have never found a NS with $M \ge 7 M_{\odot}$, as a collapsing cloud with such mass reaches BH density $\rho_{BH}$ before it reaches $\rho_{NS}$. Using more detailed considerations, the maximum is $M \le 3 M_{\odot}$ \cite{2016ARA&A..54..401O}.
By the same simple argument, we do not expect to see BH (made from the free-fall collapse of regular cold matter) which are smaller than $M \simeq 7M_{\odot}$ (in agreement with \cite{2021arXiv211103634T}), as this would require the collapse into higher densities, which violates the principles of Quantum Mechanics.

The density of a BH in Eq.\ref{eq:BHrho} is also the density of our Universe in Eq.\ref{eq:Hubble} inside $r_H$. Note from Eq.\ref{eq:Hubble} that
$H^2$ tends toward a constant $H_\Lambda^2 =   H_0^2 \Omega_\Lambda=  \Lambda/3$. As explained before, the expansion becomes asymptotically static in the physical (rest) frame with a fixed radius: $r_H \rightarrow r_\Lambda$ in Eq.\ref{eq:dS}. This $r_\Lambda$ corresponds to a future event horizon \cite{BHU1,GaztaUniverse}, like the interior of a BH. Moreover, the total mass energy inside $\rL$ is given by $ M=\rho_\Lambda V_\Lambda= r_\Lambda /2G$, which is the definition of a physical BH in Eq.\ref{eq:BHdef} for  $r_S=r_\Lambda$. Therefore, our cosmic expansion occurs inside a BH of finite size. 
For $\Omega_\Lambda  \simeq 0.7$ and $H_0 \simeq 70$ km/s/Mpc, we have:
$ M \simeq 5.5 \times 10^{22} M_{\odot}$   or $r_S \simeq 1.6 \times 10^{23}$ km. Another way to say this is that, using the FLRW metric,  at any time in cosmic history, we  can not send (received) photons to (from) distances larger than $r_\Lambda$. We are inside a finite trapped boundary, which corresponds to a BH event horizon.

We can understand this because the FLRW solution can also describe a local spherical uniform cloud of variable radius $R$ and finite mass $M$, which collapses or expands in freefall. This is a well-known concept in Newtonian gravity, which follows Eq.\ref{eq:Hubble} for arbitrary $R$. Based on Gauss’s law, each sphere $r<R$ collapses independent of what is outside $r>R$. This is also the case in General Relativity (GR), following the corollary Birkhoff’s theorem \cite{BirkhoffH}. As a consequence, the FLRW solution is also a solution for a local uniform cloud \cite{Tolman1934,Oppenheimer1939,2020PhRvD.102d4020F}.
The cloud physical radius $r=R$ follows
a boundary condition: $-2\Phi= R^2/r_H^2 = r_S/R$, which corresponds to a matching of FLRW in Eq.\ref{eq:dS} with an SW metric
$2\Phi=-r_S/r$ outside $R$ \cite{BHU1,GaztaUniverse}:
\beq
R=[r_H^2 r_S]^{1/3}.
\label{eq:R}
\eeq
This requires that either $r_H>R>r_S$
or $r_H<R<r_S$.
For a regular star, we have $R>r_S$, which implies $R<r_H$: all scales are subhorizon, as expected.
However, as discussed in Fig. \ref{fig:CMBhorizons}, in our Universe, we observe super-horizon scales $R>r_H$, which necessarily implies that $R<r_S$. Again showing that we are inside our own BH event horizon, as indicated by Fig. \ref{fig:FLRWcloud}.

\subsection{The Big Bounce}

\begin{figure}
\centering\includegraphics[width=1.\linewidth]{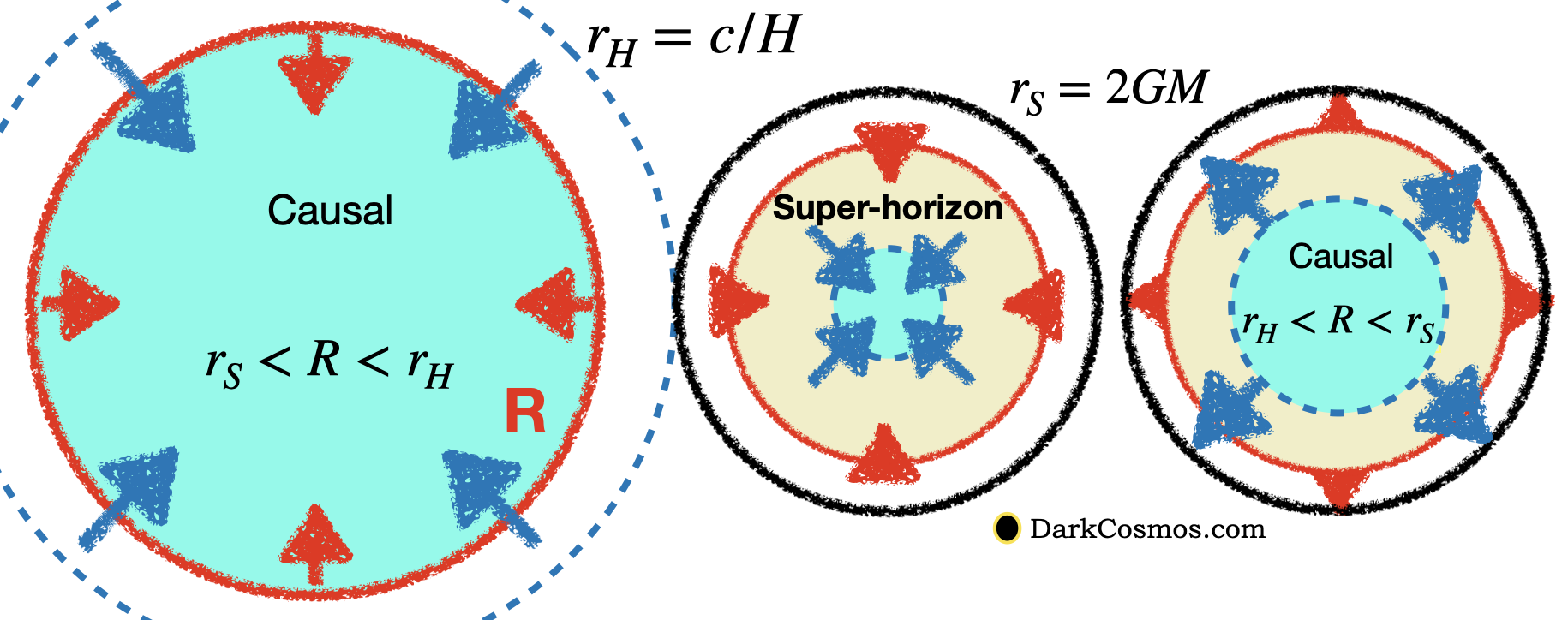}
\caption{A uniform cloud of fixed mass $M$ and size $R=(r_H^2r_S)^{1/3}$ (red circle) collapses (left) to form a BH (middle) and bounces into expansion (right) inside $r_S=2GM$ (black circle). The Hubble Horizon $r_H$ (blue dashed) moves faster than $R$, so that radial perturbations become super-horizon (yellow region) during collapse and re-enter $r_H$ during the expansion, solving the horizon problem without Inflation.}
\label{fig:FLRWcloud}
\end{figure}

Before it became a BH, the density of our FLRW cloud was so small that no interactions other than gravity could occur. Radiation escapes the cloud, so that $p=0$ ($\omega=0$). Radial co-moving shells of matter are in free-fall (time-like geodesics of constant $\chi$) and continuously pass $R=r_S$ inside its own BH horizon, as illustrated in Fig. \ref{fig:FLRWcloud}.

Solving Eq.\ref{eq:Hubble} for a collapsing dust cloud, $H \propto -a^{-3/2}$, we find that the BH forms at time:
$\tau_{BH} =-2r_S/3 \simeq -11 \textrm{Gyrs}$
before $\tau=0$ (the Big Bang) or 25 Gyr ago. The collapse continues inside until it reaches nuclear saturation (GeV) in Eq.\ref{eq:NS} and the situation is similar to the interior of a collapsing star ($R$ contains $10^{22} M_{\odot}$, but $r_H$ only has a few $M_{\odot}$). We conjecture that this leads to a Big Bounce because of Pauli´s Exclusion Principle. The collapse is halted, causing the implosion to rebound \cite{1979ARA&A..17..415B} and expand.

In the standard \LCDM model, it is speculated that reheating after inflation produces the right number of baryons per photon ($\eta = n_B/n_\gamma \simeq 6 \times 10^{-10}$) so that nucleosynthesis generates the observed primordial element abundance  \cite{2016RvMP...88a5004C}. But note that this is not a prediction, but a free parameter of the model.
Here we speculate instead that the same $\eta$ is the result of
a core collapse explosion and bounce in the BHU. The bounce happens at nuclear saturation in  Eq.\ref{eq:NS}, which corresponds to GeV$^4$ densities, well before nucleosynthesis, at MeV$^4$ energy-density (the scale factor $a$ is $10^{-3}$ times smaller). 
The Hubble radius corresponding to Eq.\ref{eq:NS} is only few km and contains a few solar masses. So the collapse mass and scale is similar to that the interior of a regular collapsing star.  Neutron density is the highest cold baryon density observed in nature. Higher densities can not be reached because of Pauli Exclusion Principle. 
This indicates that the collapse must be halted by neutron degeneracy pressure, causing the implosion to rebound \cite{1979ARA&A..17..415B}.

The different Hubble size regions explode approximately in sync at different locations because (ignoring small scale fluctuations) the background density is the same everywhere in the FLRW cloud. The collapse  energy ($H<0$) bounces into expansion energy ($H>0$).
Radiation, baryons, primordial Neutron Stars or small primordial Black Holes (PBHs) could result from each Hubble size region as compact remnants that can make up all or part of the Dark Matter $\Omega_m$. Such compact remnants do not necessarily disrupt nucleosynthesis or CMB recombination, as long as they are not too large \cite{2020ARNPS..70..355C,2022arXiv220308967B}. The small observed values of $\eta$ (inferred from element abundance) and large observed ratios for $\Omega_m/\Omega_B \simeq 4$ (inferred from CMB and BAO measurements) indicate that most remnants are in the form of radiation and compact objects (DM or $\Omega_m$) and only smaller fraction is ejected as diffused (baryonic $\Omega_B$) matter.

\begin{table*}
	\centering
	\caption{Model comparison. Observations that require explanation.} 
	\label{tab:comparison}
	\begin{tabular}{ccc} 
Cosmic observation & Big Bang (\LCDM)  & BHU \\ \hline
Expansion law & FLRW metric & FLRW metric \\
Element abundance &   Nucleosynthesis  & Nucleosynthesis \\
 baryons per foton $n_B/n_\gamma \simeq 6 \times 10^{-10}$ &  free parameter of reheating  &  Collapse remnants
 \\

CMB &  recombination  &  recombination  \\
CMB &  recombination  &  recombination  \\
All sky uniformity \& LSS seeds  & Inflation & Big Bounce \\
Cosmic acceleration $\omega = -1.03 \pm 0.03$  & Dark Energy  & BH event horizon size $r_S$\\
Age of the Universe: 14Gyr  & Dark Energy  & BH event horizon size $r_S$\\
Rot. curves, grav. lensing  \& vel. flows  & Dark Matter  &  compact stellar remnants \\
CMB fluctuations $\delta T= 10^{-5}$ &  Inflation: free parameter & perturbations during collapse/bounce  \\ 
CMB spectral index $n_S= 0.96 \pm 0.01$ &  Inflation: $1-\epsilon$  & perturbations during collapse/bounce  \\ 
$\Omega_m/\Omega_B \simeq 4$ & free parameter & compact to difuse remnants \\
$\Omega_\Lambda/\Omega_m \simeq 3$ & free parameter & time to de-Sitter phase \\
Large scales CMB anomalies & Cosmic Variance (bad luck) & super-horizon cut-off Eq.\ref{eq:cut-off} \\
Hubble tension & Systematic effects &  super-horizon perturbations \\
flat universe $k=0$  & Inflation & topology of empty space 
\\
	\end{tabular}
\end{table*}

\subsection{Trapped Inside a Black Hole}

Because $R \propto r_H^{2/3}$ in Eq. \ref{eq:R}, $r_H \propto \tau$ always expands or collapses faster than $R$. Once inside $r_S$,
perturbations of size $r$ exit the horizon $r>r_H$ during collapse
and re-enter $r_H>r$ as the expansion occurs (see Fig. \ref{fig:FLRWcloud}), solving the horizon problem without Inflation.
Once the FLRW cloud becomes a BH, no events can escape $r_S$. This translates into an expansion that freezes out or becomes static in physical units when $R$ approaches $r_S$. 
 According to Eq.~(\ref{eq:dS}) this is observed as cosmic acceleration for a comoving observer:
$\dot{H}=0$ or $q=1$. Thus, $r_S$ behaves like a $\Lambda$ term ($\Lambda=3r_S^{-2}$), despite our choice of $\Lambda=0$ in the background where the collapse occurred \cite{BHU1}.

Such a $r_S$ boundary imposes a cut-off in the spectrum of super-horizon perturbations generated during collapse and bounce. During the collapse, $H \propto -a^{-3/2}$ ($\Lambda=k=p=0$), the maximum radial comovil separation (and corresponding maximum transverse CMB angle) between two events is:
\beq
\chi_{max} = \int_{r_S}^0 \frac{c da}{H a^2} = 2 r_S  \,\,\, ; \,\,\,
\theta_{max} = \pi r_S/\chi_*,
\label{eq:cut-off}
\eeq
where $\chi_*$ is the angular diameter distance to the CMB
estimated with the type Ia SN
 value of $H_0$ with $\Omega_m \simeq 0.25 \pm 0.05$ and $k=0$ (i.e., an apparent $\Omega_\Lambda \simeq 0.75$), as measured today locally. Note that $\chi_{max}$ scales as $H_0^{-1}$. The good agreement of these predictions with the data in Fig. \ref{fig:CMBhorizons2} indicates that cosmic expansion originates from a gravitational collapse and bounce inside $r_S$ with $\Lambda=0$. Models of cosmic inflation with a cut-off or a phase transition \cite{Barriga} still need $\Lambda$ to account for cosmic acceleration.

\section{Discussion and Conclusion}
\label{sec:conclusion}

The large-scale structures that we observe in Cosmic maps are measured to be adiabatic and scale invariant, as predicted by Harrison-Zeldovich-Peebles \cite{Zeldovich1970,Harrison1970,PeeblesYu}, long before Inflation. The seeds could originate during the BHU collapsing phase or during Inflation. However, we find that, at the largest scales, structures are neither scale invariant nor adiabatic:
\begin{itemize}
    \item there is a cut-off scale in the largest super-horizon modes as measured in $\delta T$ (Fig.\ref{fig:CMBhorizons2}). This is reflected in the homogeneity scale $\theta_{\mathcal{H}} = 65.9 \pm 9.2$ deg.,
    \item a similar cut-off scale is measured in the values of $H_0$ (or other cosmological parameters) fit at different positions in the CMB sky (Fig.\ref{fig:CMBhorizons}),
    \item the amplitude of these super-horizon modes in radiation ($\delta_T \simeq 10^{-5}$) is much smaller than those measured in $H_0$ or $\rho$: $\delta_\Phi \simeq 0.2$ in Eq.\ref{eq:deltaH0}, both within the CMB and between the CMB and SNe,
    \item the location of these super-horizon modes in radiation do not trace well the ones in $\delta_\Phi$ as shown by the comparison of top and bottom panels in Fig.\ref{fig:CMBhorizons} or in Fig.\ref{fig:CMBhorizons3}.
\end{itemize}
Fig.\ref{fig:CMBhorizons2} shows good agreement of these observations with the prediction of Eq.\ref{eq:cut-off}. 
This supports the interpretation that the observed expansion is bounded by $r_S$, which is the source of cosmic acceleration and provides an interpretation for $\Lambda=3/r_S^2$. 
This can not be explained by the  simplest 
models of Inflation.
It could indicate instead that cosmic expansion results from a collapse and bounce. 

The complete GR solution for such collapse is given by the BHU in 
\cite{BHU1,GaztaUniverse}, which models the gravitational collapse of a FLRW dust cloud first studied by \cite{Lemaitre,Tolman1934,Oppenheimer1939}. 
\ref{BHU} provides a brief summary of the BHU as a formal GR solution.
Note how this is quite different from the LBT model \cite{PhysRevD.83.103515}, which tries to explain cosmic acceleration with a spherical void around us.  You could picture our FLRW cloud as an LBT model with a central over-density (instead of a void)
where the outside background is empty.
But LBT provides a smooth transition between the two backgrounds while for the BHU the two backgrounds are uniformed and causally separated by an event horizon.

The idea of a bouncing universe is not new
\cite{2008PhR...463..127N,2018Ijjas}, but previous proposals do not happen inside a BH and require some form of Modified Gravity laws to avoid the initial singularity, often resulting in cyclic models. The BHU proposal only uses the known laws of  Physics:  Quantum Mechanics and General Relativity. The bounce happens at moderate energy scales (GeV), many orders of magnitude smaller than that of Inflation or Planck scales. Thus, we don't need Quantum Gravity or Inflation to understand the origin of cosmic expansion.
The collapse is halted by Quantum Mechanics. When the collapse reaches nuclear saturation in Eq.\ref{eq:NS}, it bounces back
like a core collapse supernova. Such bounce could explain the large observed metric fluctuations in Eq.\ref{eq:deltaH0}. 

This idea is speculative, the same way Inflation is speculative. The main difference is that Inflation happens at energies that we will never be able to test, whereas the core collapse bounce in the BHU can be modeled and tested using the same Nuclear Astrophysics and observations that we use to understand Neutron stars, pulsars or core collapsed supernovae.

Our FLRW cloud collapse occurred in an existing background. We do not know what else is out there or how it started, but we have assumed that it has the simplest topology: flat with $\Lambda=0$, as in Minkowski space.
The observed accelerated expansion $q$ is usually attributed to a $\Lambda$ term in the background with $\Omega_\Lambda \simeq 0.75 \pm 0.05$ in Eq. \ref{eq:Hubble}. During expansion, this is equivalent to the BHU event horizon $r_S$, where $\Lambda =3/r_S^2$ \cite{BHU1}. We can only distinguish the effects of a true $\Lambda$ from $r_S$ during the collapse. Because of the equivalence principle, the dynamics of collapsing shells in free fall are not affected by $r_S$. However, if $q$ originates from a true $\Lambda$ (rather than from $r_S$ with $\Lambda=0$), it will change the collapse time so that $\chi_{max} \simeq 3.2 r_S$ in Eq.\ref{eq:cut-off} (red lines in Fig. \ref{fig:CMBhorizons2}), which is clearly ruled out by the $H_0$ data in Fig. \ref{fig:CMBhorizons2}, which favors $\Lambda=0$.

The BH collapse time is proportional to $M$. A mass $M \simeq 5.5 \times 10^{22} M_{\odot}$ is just the right one to allow enough time for galaxies and planets to form before the de-Sitter phase dominates. This provides an anthropic explanation 
as to why we live inside such a large expanding BH \cite{BHU2}. The BHU solution can also be used to model the interior of smaller BHs, but they will not have time to form regular galaxies or stars before they reach the asymptotically static de-Sitter phase.

The fact that the universe might be generated from the inside of a BH has been studied extensively in the literature \cite{Smolin1992,Easson,Daghigh,Firouzjahi,Yokoyama,Dymnikova}. But 
most approaches involve modifications to classical GR.  Among the classical GR solutions are the Bubble or Baby Universe models, where the BH interior is de-Sitter metric \cite{Gonzalez-Diaz,1989PhLA..138...89G,1987PhRvD..35.1747B,1989PhLB..216..272F,Aguirre,gravastar2015,Garriga16,PBH3}.
The BHU solution is similar, but has some important differences. In the BHU, no surface terms (or Bubble) is needed and the matter and radiation inside are regular. There is no need for a false vacuum in the BHU. In this respect the BHU is not quite a Bubble Universe.

Our expansion will become static inside a BH in a larger and older background, possibly containing other BHUs. This provides another layer to the Copernican Principle and avoids the important 
conceptual problems of the standard Big Bang model: Inflation and Dark Energy, which are hard to falsify with observations because such ingredients are not fundamentally understood. 
They are an ad hoc proposal to address data inconsistencies, such as cosmic acceleration, structure formation or the horizon problem.
The BHU idea can be more easily tested because is a much simpler scenario: it only uses known laws of Physics and known energy and particle components.
The BHU model could be falsified for example if we measure $\omega \neq -1$. 
Further work is needed to estimate the amplitude of perturbations, composition and fraction of compact to diffuse remnants that resulted from the Big Bounce. This could potentially explain from first principles some key observations  in Table \ref{tab:comparison}, like $\eta$, $\delta T$ or $\Omega_m/\Omega_B$, which are currently modeled as free parameters in the \LCDM model.

\section*{Acknowledgements}

This work was partially supported by grants from Spain MCIN/AEI/10.13039/501100011033  grants PGC2018-102021-B-100, PID2021-128989NB-I00 and Unidad de Excelencia María de Maeztu CEX2020-001058-M and from European Union funding LACEGAL 734374 and EWC 776247. IEEC is funded by Generalitat de~Catalunya.  BCQ  acknowledge support from a PhD scholarship from the Secretaria d’Universitats i Recerca de la Generalitat de Catalunya i del Fons Social Europeu. EG thanks 
Angela Olinto and Sergio Assad for their hospitality during the summer of 2022, when the latest version of this paper together with \cite{BHU1,BHU2} were completed, extracted from earlier unpublished drafts \cite{hal-03106344,hal-03344159}, which preceded later review articles on the same topic \cite{GaztaUniverse,Gazta2022sym}

\appendix

\section{The Black Hole Universe (BHU)}
\label{BHU}

The Friedmann–Lemaitre–Robertson–Walker (FLRW) flat metric in comoving coordinates $\xi^\alpha=(\tau,\chi,\theta,\delta)$, corresponds to an homogeneous and isotropic space:
\beq
ds^2 =f_{\alpha\beta} d\xi^\alpha d\xi^\beta = -d\tau^2 + a(\tau)^2\left[ d\chi^2 + \chi^2 \dA \right],
\label{eq:frw}
\eeq
For a perfect fluid with density $\rho$ and pressure $p$, the solution to the field equations in a flat space is the well known
 Eq.\ref{eq:Hubble}.
 
We will show here that the FLRW solution to GR  also works for a finite spherical volume $r=a\chi<R$. We call this solution the FLRW cloud. When the FLRW cloud is
inside its SW radius $R<r_S$, we call this a BHU. Recall that $r_S$ plays the same role as a $\Lambda$ term with $\Lambda=3/r_S^2$ in our expanding Universe. But such effective $\Lambda$ just corresponds to the total energy-mass $M$ inside $R$. Because we have recently measured an effective $\Lambda$, we now know that our FLRW cloud is inside $r_S$, so we are inside a BHU.

There are several ways to show that the FLRW cloud or BHU are exact solutions to GR equations. Here we present three alternative versions. First the one based on  known classical solutions, second the one based on the frame duality and finally one using junction conditions.

\subsection{Classical solutions}

The first evidence for the FLRW cloud comes from the original expanding universe discovery presented by Lemaitre \cite{Lemaitre}. This was further explored by 
Tolman \cite{Tolman1934} and Oppenheimer \& Snyder \cite{Oppenheimer1939}. As detailed in Tolman \cite{Tolman1934} in his Application e), a combination of different FLRW 
distributions (with different densities at different radius) is also a solution to GR field equations. This is a consequence of the
corollary Birkhoff’s theorem \cite{BirkhoffH} for spherically symmetric solutions, as each sphere $r<R$ evolves with independence of what is outside $r>R$. This solution have also been generalized by Vaidya \cite{Vaidya1968} for the interior of Schwarzschild metric $r<r_S$ and found the particular cases of the FLRW and the Schwarzschild solutions.
Another example is McVittie's metric \cite{Kaloper2010}, which corresponds to a junction of a exterior FLRW metric with the interior Schwarzschild solutions.

\subsection{Frame duality}

Consider the most general form of a
metric with spherical symmetry in physical or Schwarzschild (SW) coordinates $x^\mu=(t,r,\theta, \varphi)$~\cite{Padmanabhan,Dodelson}:
\beq
 ds^2 = g_{\mu\nu} dx^\mu dx^\nu =
 -(1+2\Psi) dt^2 +  \frac{dr^2}{1+2\Phi} + r^2 \dA^2,
\label{eq:newFRW}
 \eeq 
 where  $\Phi=\Phi(t,r)$ and $\Psi=\Phi(t,r)$ are the generic gravitational potentials.
  The Weyl potential $\Phi_W$ is the geometric mean of the two:
 \beq
 ( 1+2\Phi_W)^2= (1+2\Phi) (1+2\Psi).
 \eeq
$\Psi$ describes propagation of non-relativist particles and $\Phi_W$ the propagation of light. The solution to Einstein's field equations for empty space ($\rho=p=\rho_\Lambda=0$) results in the well known Schwarzschild (SW) metric:
 \beq
2\Phi= 2\Psi= - 2GM/r \equiv -  r_{S}/r,
\label{eq:Schwarzschild}
 \eeq 
 which describes a singular BH  of mass $M$ at $r=0$.
The FLRW metric is also spherically symmetric, so it must be a particular case of the metric Eq.\ref{eq:newFRW}. 
There must exist a change of variables $C$ from  $x^\mu= [t, r]$ to comoving coordinates $\xi^\nu=[\tau,\chi]$, where $r=a(\tau) \chi$ and angular variables $(\theta, \delta)$ remain the same, such that:
\beq
C^T
\begin{pmatrix} 
-(1+2\Psi)
& 0 
 \\ 
0 & (1+2\Phi)^{-1} 
\end{pmatrix}
C
=
\begin{pmatrix} 
-1 & 0 
 \\ 
0 & a^2
\end{pmatrix}.
\label{eq:fab}
\eeq
The unique solution is \cite{BHU1}:
\beq
C \equiv \begin{pmatrix}
  \partial_\tau t &  \partial_\chi t \\
  \partial_\tau r    &  \partial_\chi r  \\
 \end{pmatrix}
=
\begin{pmatrix} 
(1+2\Phi_W)^{-1} & a r H (1+2\Phi_W)^{-1}
 \\ 
r H  &  a
\end{pmatrix},
\label{eq:xi2xH}
\eeq
where $2\Phi= -r^2 H^2(\tau)$ while  $a(\tau)$ and $\Psi$ are arbitrary. This reproduces Eq.\ref{eq:dS} which we  found using a Lorentz transformation and is what we call the dual frame representation of the FLRW metric. This is a particular case of the change of variables between section I and section II in \cite{Oppenheimer1939}.
We can now find the solution in SW coordinates for the case where we
have uniform density $\rho$ inside some radius $R$ and empty space outside:
\bea
\rho(t,r) &=& \left\{ \begin{array}{ll} 
0  &  {\text{for}} ~ ~ r>R \\
\rho & {\text{for}} ~ ~r<R \\
\end{array} \right. .
\label{eq:rhoVu}
\eea
 Because we have spherical symmetry we can  apply Gauss law (or Birkhoff Theorem) to both sides of $R$.  For $r>R$ the solution is clearly SW metric. For $r<R$ the solution is FLRW, which in the SW frame  we have just show it corresponds to $2\Phi=-r^2 H^2$. So the solution for $\Phi$ reads:
\beq
-2\Phi(t,r) = \left\{ \begin{array}{ll} 
r_{S}/r  &  {\text{for}} ~ ~ r>R\\
 r^2 H^2   & {\text{for}} ~ ~r<R \\
\end{array} \right. .
\label{eq:BH.u2}
\eeq
where $H^2= \frac{8\pi G}{3} \rho$ is given by the FLRW solution of Eq.\ref{eq:Hubble}.
At the junction $r=R$ this solution reproduces Eq.\ref{eq:R}.
To find $\Psi$ and $t=t(\tau,\chi)$, we need to integrate Eq.\ref{eq:xi2xH} with $2\Phi= -r^2 H^2(\tau)$, for a given $H(\tau)$ solution of Eq.\ref{eq:Hubble}.
For example, for $H(\tau)=H_{\Lambda}$ the solution is $\Psi=\Phi$ and 
\beq
t = t(\tau,\chi)= \tau - \frac{1}{2H_\Lambda} 
\ln{[1-\chi^2 a^2 H^2_\Lambda]},
\label{eq:Lanczos}
\eeq
which reproduces the static de-Sitter metric: $2\Phi=-r^2 H_\Lambda^2$.

\subsection{Junction conditions}

We can arrive at the same FLRW cloud (or BHU) solution using Israel's  junction conditions (~\cite{Israel,Barrabes1991}). Here we just summarize the calculations previously presented in \cite{BHU1}.
The FLRW solution will be smoothly joined with the Schwarzschild solution at some 
hypersurface junction $\Sigma$, 
which will be given by $r=R$ as in Eq.\ref{eq:rhoVu}. The junction conditions require that the metric and its derivative (the extrinsic curvature $K$) on both sides of the junction are equal. The join metric then provides a new solution to GR.
 
Consider the case where $R$ corresponds to a fix
comoving coordinate $\chiSW$ so that $R= a(\tau) \chiSW$. The only free variable remaining is $\tau$,  the FLRW comoving time 
(the solid angle $d\Omega$ is the same in both metrics as we have spherical symmetry).
This corresponds to a freefall timelike geodesic spherical surface with fixed mass $M$ inside $r<R$.
The induced 3D metric on $\Sigma$ is $h_{\alpha\beta}^\in$ has
coordinates $dy^\alpha=(d\tau,d\delta,d\theta)$:
which just corresponds to $d\chi=0$ in Eq.\ref{eq:frw}:
\beq
ds^2_{\Sigma^\in}= h_{\alpha\beta}^\in  dy^\alpha dy^\beta= -d\tau^2 + a^2(\tau) \chiSW ^2 \dA^2.
\label{eq:Sigma}
\eeq
For the outside SW frame, the same junction $\Sigma^{\out}$ is described by some unknown functions $r=R(\tau)$ and $t=T(\tau)$, where $t$ and $r$ are the time and radial coordinates in the physical SW frame of Eq.\ref{eq:newFRW}. We then have:
\beq
dr= \dot{R} d\tau ~~;~~ dt= \dot{T} d\tau,
\label{eq:RT}
\eeq
where the dot refers to derivatives with respect to $\tau$.  The induced metric $h^{\out}$ estimated from the outside 
SW metric Eq.\ref{eq:Schwarzschild} becomes:
\bea
ds^2_{\Sigma^\out} &=& h_{\alpha\beta}^\out  dy^\alpha dy^\beta =
-F dt^2 + \frac{dr^2}{F} + r^2  \dA^2   \nonumber \\
&=& - (F\dot{T}^2- \dot{R}^2/F) d\tau^2 + R^2 \dA^2,  \label{eq:dS+}
\eea
where $F \equiv 1- r_{S}/R$. Comparing Equation~(\ref{eq:Sigma})
with Equation~(\ref{eq:dS+}), the first matching
conditions $h^{\in}=h^{\out}$ are then:
\beq
R(\tau) = a(\tau) \chiSW ~~;~~   F\dot{T} =  \sqrt{\dot{R}^2+F} \equiv \beta({R,\dot{R}}).
\label{eq:matching}
\eeq
{For} any given $a(\tau)$ and $\chiSW$ we can find both $R(\tau)$ and $\beta(\tau)$. 
We also want the derivative of the metric to be continuous at $\Sigma$. For this, we  estimate the extrinsic curvature $K^\pm$ normal to $\Sigma^\pm$ from each side of the hypersurface:
\beq
K_{\alpha\beta} = 
-\left[ \partial_a n_b - n_c \Gamma^c_{ab} \right]  e^a_\alpha e^b_\beta, 
\label{eq:Kab}
\eeq
where $e^a_\alpha =\partial x^a/\partial y^\alpha$ and $n_a$ is the 4D vector normal to $\Sigma$.
The outward 4D velocity is $u^a = e_\tau^a = (1,0,0,0)$ and
the normal to $\Sigma^\in$ on the inside is then 
$n^\in= (0,  a,0,0)$.
On the outside  $u^a = (\dot{T},\dot{R},0,0)$ and
$n^\out =(-\dot{R},\dot{T},0,0)$. It is straightforward to verify that: $n_a u^a =0$ and $n_a n^a =+1$ (for a timelike surface) for both $n^\in$ and $n^\out$. 
We then find that the extrinsic curvature in Equation~(\ref{eq:Kab}) to the $\Sigma$ junction, estimated with the inside FLRW metric, i.e., $K^\in$ is:
\bea
K^\in_{\tau \tau} &=&  -(\partial_\tau n^\in_\tau -  a \Gamma_{\tau\tau}^{\chi}) e_\tau^\tau e_\tau^\tau =0    \nonumber \\
K^\in_{\theta \theta} &=&     a \Gamma_{\theta\theta}^{\chi}  e_\theta^\theta e_\theta^\theta= -a \chi_* = -R. 
\label{eq:Kin}
\eea
{For} the SW metric:
\bea
K^\out_{\tau\tau}   &=&  \ddot{R} \dot{T} - \dot{R} \ddot{T} +\frac{\dot{T} r_{S}}{2 R^2 F} (\dot{T}^2 F^2-3 \dot{R}^2) =
\frac{\dot{\beta}}{\dot{R}}  \nonumber  \\ 
 K^\out_{\theta\theta} &=&   \dot{T} \Gamma_{\theta\theta}^r =  -\dot{T} F R =  -\beta R, 
\label{eq:Kout}
\eea
where we have used the definition of $\beta$ in Equation~(\ref{eq:matching}). In both cases
$K_{\delta\delta} = \sin^2{\theta} K_{\theta\theta}$, so that when $K_{\theta\theta}^\in = K_{\theta\theta}^\out$,
it follows that  $K_{\delta\delta}^\in = K_{\delta\delta}^\out$.
Comparing Equation~(\ref{eq:Kin}) with Equation~(\ref{eq:Kout}), the second matching conditions $K^{\in}=K^{\out}$ require $\beta=1$, which using Equation~(\ref{eq:matching}) 
reproduces again the junction in 
Eq.\ref{eq:R}. So the two metrics and  derivatives (the extrinsic curvature) are identical in the hypersurface defined by $r=R$ as long as $R$ follows Eq.\ref{eq:R}. This completes the proof that the FLRW cloud is an exact solution of GR without surface terms. For more details see~\cite{BHU1}.

\black
 \bibliographystyle{elsarticle-num} 
 \bibliography{Singular}





\end{document}